\documentclass[conference]{IEEEtran}

\usepackage{cite}
\usepackage{amsmath,amssymb,amsthm}
\usepackage{graphicx}
\usepackage{booktabs}
\usepackage{xcolor}
\usepackage{url}
\usepackage[hidelinks]{hyperref}

\newtheorem{proposition}{Proposition}

\newcommand{\PARTpcdsHardRandomOracleMax}{2.99}
\newcommand{\PARTpcdsHardRandomOracleMin}{2.52}

\newcommand{\PARTpchHardRandomOracleMax}{2.11}
\newcommand{\PARTpchHardRandomOracleMin}{1.76}

\newcommand{\PATpchProcessorsRandomAffineParallelOneZero}{1.28}
\newcommand{\PATpchProcessorsRandomLowdimTemplateOneZero}{1.23}

\newcommand{\PATpchProcessorsRandomSpeedOneZero}{1.41}

\newcommand{\PAffTPCDSAbove}{7}
\newcommand{\PAffTPCDSBelow}{10}
\newcommand{\PAffTPCDSPairs}{30}
\newcommand{\PAffTPCHAbove}{0}
\newcommand{\PAffTPCHBelow}{20}
\newcommand{\PAffTPCHPairs}{30}
\newcommand{\PHHFiveN}{18}
\newcommand{\PHHFiveOk}{18}
\newcommand{\PHHFiveOkCI}{16}
\newcommand{\PHHFivePass}{holds}

\newcommand{\PHHFourMax}{1.08}
\newcommand{\PHHFourMaxAll}{1.12}

\newcommand{\PHHFourPass}{holds}
\newcommand{\PHHOneMin}{0.74}

\newcommand{\PHHOnePass}{holds}
\newcommand{\PHHThreeN}{12}
\newcommand{\PHHThreeOk}{10}
\newcommand{\PHHThreeOkCI}{8}
\newcommand{\PHHThreePass}{holds}
\newcommand{\PHHTwoN}{18}
\newcommand{\PHHTwoPass}{holds}
\newcommand{\PHHTwoVsRidgeOk}{16}
\newcommand{\PHHTwoVsSpeedOk}{18}
\newcommand{\PHPairs}{18}
\newcommand{\PLTpcdsKOneZeroZeroDOptimalCalibrationS}{187.7}

\newcommand{\PLTpcdsKOneZeroZeroKCenterOneNn}{1.18}

\newcommand{\PLTpcdsKOneZeroZeroRandomLowdimTemplate}{1.23}

\newcommand{\PLTpcdsKOneZeroZeroRoundRobinCalibrationS}{109.9}
\newcommand{\PLTpcdsKOneZeroZeroRoundRobinLowdimTemplate}{1.19}

\newcommand{\PLTpcdsKOneZeroZeroRrCheapCalibrationS}{87.9}

\newcommand{\PLTpchKTwoFiveDOptimalCalibrationS}{15.7}

\newcommand{\PLTpchKTwoFiveKCenterOneNn}{1.13}

\newcommand{\PLTpchKTwoFiveRandomLowdimTemplate}{1.17}

\newcommand{\PLTpchKTwoFiveRoundRobinCalibrationS}{23.6}
\newcommand{\PLTpchKTwoFiveRoundRobinLowdimTemplate}{1.10}

\newcommand{\PLTpchKTwoFiveRrCheapCalibrationS}{16.4}

\newcommand{\PLadTpcdsForestResidRandom}{1.28}
\newcommand{\PLadTpcdsLookupRoundRobin}{1.24}
\newcommand{\PLadTpcdsLowdimTemplateControlRoundRobin}{1.18}
\newcommand{\PLadTpcdsLowdimTemplateRoundRobin}{1.19}

\newcommand{\PLadTpcdsOracleRandom}{1.26}
\newcommand{\PLadTpcdsPooledRandom}{1.36}

\newcommand{\PLadTpcdsTargetRandom}{1.80}
\newcommand{\PLadTpcdsTemplateRawRoundRobin}{1.19}
\newcommand{\PLadTpchForestResidRandom}{1.22}
\newcommand{\PLadTpchLookupRoundRobin}{1.24}
\newcommand{\PLadTpchLowdimTemplateControlRoundRobin}{1.08}
\newcommand{\PLadTpchLowdimTemplateRoundRobin}{1.10}

\newcommand{\PLadTpchOracleRandom}{1.05}
\newcommand{\PLadTpchPooledRandom}{1.36}

\newcommand{\PLadTpchTargetRandom}{1.51}
\newcommand{\PLadTpchTemplateRawRoundRobin}{1.12}
\newcommand{\PNnBelowPrimTPCDS}{22}

\newcommand{\PNoneTpcdsKOneZeroZero}{1.96}

\newcommand{\PNoneTpchKTwoFive}{1.83}
\newcommand{\PPrimOracleTPCDSMax}{0.95}
\newcommand{\PPrimOracleTPCDSMin}{0.93}

\newcommand{\PSRTpcdsEpsilonTwoMax}{0.93}

\newcommand{\PSRTpcdsNoiseMax}{0.22}

\newcommand{\PSRTpcdsShareMax}{0.94}

\newcommand{\PSRTpchEpsilonTwoMin}{0.68}
\newcommand{\PSRTpchNoiseMax}{0.08}

\newcommand{\PSRTpchRTwoLotMax}{0.53}

\newcommand{\PSRTpchShareMin}{0.69}

\newcommand{\PSTpcdsMOnetoMFiveRTwoBoth}{0.02}

\newcommand{\PSTpcdsMOnetoMFiveSlowdown}{1.03}

\newcommand{\PSTpcdsMOnetoMThreeRTwoSize}{0.64}

\newcommand{\PSTpcdsMOnetoMTwoRTwoParallel}{0.44}

\newcommand{\PSTpcdsMThreetoMFourParallelX}{2.05}

\newcommand{\PSTpcdsMThreetoMFourSerial}{0.92}

\newcommand{\PSTpcdsMThreetoMSixParallelX}{1.25}

\newcommand{\PSTpcdsMTwotoMFourRTwoSize}{0.41}

\newcommand{\PSTpchMOnetoMFiveRTwoBoth}{0.08}

\newcommand{\PSTpchMOnetoMFiveSerial}{1.47}

\newcommand{\PSTpchMOnetoMFiveSlowdown}{1.08}

\newcommand{\PSTpchMOnetoMThreeRTwoSize}{0.52}

\newcommand{\PSTpchMOnetoMTwoParallelX}{2.66}

\newcommand{\PSTpchMOnetoMTwoRTwoParallel}{0.61}

\newcommand{\PSTpchMOnetoMTwoSerial}{1.12}

\newcommand{\PSTpchMThreetoMSixParallelX}{1.32}

\newcommand{\PSTpchMTwotoMFourRTwoSize}{0.22}

\newcommand{\PSpeedTpcdsKFive}{1.50}

\newcommand{\PSpeedTpcdsKTwoFive}{1.45}
\newcommand{\PSpeedTpchKFive}{1.39}

\newcommand{\PSpeedTpchKTwoFive}{1.37}
\newcommand{\PSpreadTPCDS}{1.09}

\newcommand{\OpHwRange}{1.0 to 2.0}
\newcommand{\OpMemIndex}{1.3 to 1.9}
\newcommand{\OpMemSeq}{1.0 to 1.3}
\newcommand{\OpProcParallel}{2.0 to 3.2}

\newcommand{\OpServerProcParallel}{2.0 to 3.0}

\newcommand{\SrcTpcdsCostOracle}{2.61}

\newcommand{\SrcTpcdsCostPrimary}{1.19}

\newcommand{\SrcTpchCostOracle}{1.87}

\newcommand{\SrcTpchCostPrimary}{1.15}

\newcommand{\SrcWinsTotal}{240}

\newcommand{\StreamFinalLowdimTemplateLastFiveZero}{1.15}
\newcommand{\StreamFinalNever}{1.91}
\newcommand{\StreamFinalRidgeAll}{1.49}
\newcommand{\StreamFinalRidgeLastFiveZero}{1.21}

\newcommand{\SingleTpcdsFirstControl}{1.24}
\newcommand{\SingleTpcdsFirstLowdimTemplate}{1.25}

\newcommand{\SingleTpcdsWarmLowdimTemplate}{1.22}

\newcommand{\SingleTpchFirstControl}{1.12}
\newcommand{\SingleTpchFirstLowdimTemplate}{1.14}

\newcommand{\SingleTpchWarmLowdimTemplate}{1.12}

\newcommand{\NNDevEpochs}{800}
\newcommand{\NNDevLr}{0.003}
\newcommand{\NNDevHidden}{128}

\newcommand{\NNDevDropout}{0}
\newcommand{\NNDevSourceSeconds}{59}
\newcommand{\NNDevTpchParams}{19,732}

\newcommand{\NNDevTpchRrHybNoneKOneZero}{1.83}

\newcommand{\NNDevTpchRrHybTplKOneZero}{1.22}

\newcommand{\NNDevTpchRrHybTplKOneZeroZero}{1.08}

\newcommand{\NNDevTpchRrNnNoneKOneZero}{1.84}

\newcommand{\NNDevTpchRrNnFullKOneZero}{1.28}

\newcommand{\NNDevTpchRrNnFullKOneZeroZero}{1.06}

\newcommand{\NNDevTpchHSixFullKOneZeroPoint}{29}

\newcommand{\NNDevTpchHSixFullKOneZeroZeroBelow}{0}

\newcommand{\NNDevTpchHSixFullKOneZeroZeroMaxHi}{1.082}

\newcommand{\NNDevTpcdsParams}{25,902}

\newcommand{\NNDevTpcdsRrHybNoneKOneZero}{1.96}

\newcommand{\NNDevTpcdsRrHybTplKOneZero}{1.42}

\newcommand{\NNDevTpcdsRrHybTplKOneZeroZero}{1.19}

\newcommand{\NNDevTpcdsRrNnNoneKOneZero}{1.89}

\newcommand{\NNDevTpcdsRrNnTplKOneZeroZero}{1.17}

\newcommand{\NNDevTpcdsRrNnOracleKOneZeroZero}{1.16}

\newcommand{\NNDevTpcdsRrNnFullKOneZero}{1.46}

\newcommand{\NNDevTpcdsRrNnFullKOneZeroZero}{1.17}

\newcommand{\NNDevTpcdsHSixFullKOneZeroPoint}{19}

\newcommand{\NNDevTpcdsHSixFullKOneZeroZeroBelow}{0}

\newcommand{\NNDevTpcdsHSixFullKOneZeroZeroMaxHi}{1.112}

\newcommand{\EnvNoiseMedQMin}{1.03}
\newcommand{\EnvNoiseMedQMax}{1.09}

\newcommand{\EnvPassTpchMin}{21}
\newcommand{\EnvPassTpchMax}{75}

\newcommand{\XtraAnalogPairs}{8}
\newcommand{\XtraTpchTransferKFive}{1.18}
\newcommand{\XtraTpchPrimaryAnalogKFive}{1.26}
\newcommand{\XtraTpchTransferWinsKFive}{8}

\newcommand{\XtraTpchTransferWinsKOneZero}{6}
\newcommand{\XtraTpchTransferKTwoFive}{1.16}
\newcommand{\XtraTpchPrimaryAnalogKTwoFive}{1.09}

\newcommand{\XtraTpchForestResidRandKTwoZeroZero}{1.06}
\newcommand{\XtraTpchTargetRandKTwoZeroZero}{1.07}

\newcommand{\XtraTpchPrimaryRandKTwoZeroZero}{1.07}

\newcommand{\XtraTpchForestTplRrKTwoFive}{1.20}
\newcommand{\XtraTpchPrimaryRrKTwoFive}{1.10}
\newcommand{\XtraTpchForestTplWinsRrKTwoFive}{0}

\newcommand{\XtraTpcdsTransferKFive}{1.42}
\newcommand{\XtraTpcdsPrimaryAnalogKFive}{1.42}

\newcommand{\XtraTpcdsForestTplRrKOneZeroZero}{1.32}
\newcommand{\XtraTpcdsPrimaryRrKOneZeroZero}{1.19}
\newcommand{\XtraTpcdsForestTplWinsRrKOneZeroZero}{0}

\newcommand{\XtraHardTransferMin}{0.99}
\newcommand{\XtraHardTransferMax}{1.08}
\newcommand{\XtraHardBestPct}{2}
\newcommand{\XtraNnGainPct}{2}
\newcommand{\XtraTpchHThreeCI}{4}
\newcommand{\XtraTpchHThreeN}{4}
\newcommand{\XtraTpcdsHThreeCI}{4}
\newcommand{\XtraTpcdsHThreeN}{8}

\newcommand{\XhoTpchInPrimary}{1.10}
\newcommand{\XhoTpchInControl}{1.09}

\newcommand{\XhoTpchOutPrimary}{1.13}
\newcommand{\XhoTpchOutControl}{1.12}

\newcommand{\XhoTpcdsInPrimary}{1.21}
\newcommand{\XhoTpcdsInControl}{1.20}

\newcommand{\XhoTpcdsOutPrimary}{1.18}
\newcommand{\XhoTpcdsOutControl}{1.17}

\newcommand{\XtlTpchPNinetySpeed}{1.87}

\newcommand{\XtlTpchBiasSpeed}{1.11}
\newcommand{\XtlTpchPNinetyPrimary}{1.39}

\newcommand{\XtlTpchBiasPrimary}{1.00}
\newcommand{\XtlTpchPNinetyOracle}{1.18}

\newcommand{\XtlTpcdsPNinetySpeed}{2.03}

\newcommand{\XtlTpcdsBiasSpeed}{1.10}
\newcommand{\XtlTpcdsPNinetyPrimary}{1.42}
\newcommand{\XtlTpcdsPNinetyNinePrimary}{6.69}
\newcommand{\XtlTpcdsBiasPrimary}{1.00}
\newcommand{\XtlTpcdsPNinetyOracle}{1.60}
\newcommand{\XtlTpcdsPNinetyNineOracle}{7.41}

\newcommand{\XcNewJobHardBestPct}{5}
\newcommand{\XcNewJobHardOracleMax}{1.01}
\newcommand{\XcNewJobHardOracleMin}{0.88}

\newcommand{\XcNewJobShrinkParallelMax}{2.61}
\newcommand{\XcNewJobShrinkParallelMin}{1.43}
\newcommand{\XcNewJobShrinkSerialMax}{1.01}
\newcommand{\XcNewJobShrinkSerialMin}{0.97}
\newcommand{\XcNewTpchTenHardBestPct}{3}
\newcommand{\XcNewTpchTenHardOracleMax}{1.03}
\newcommand{\XcNewTpchTenHardOracleMin}{0.93}

\newcommand{\XcNewTpchTenShrinkParallelMax}{2.42}
\newcommand{\XcNewTpchTenShrinkParallelMin}{1.50}
\newcommand{\XcNewTpchTenShrinkSerialMax}{0.98}
\newcommand{\XcNewTpchTenShrinkSerialMin}{0.94}

\newcommand{\XcOneTpchTenTpchKOneZeroAffinePar}{1.18}

\newcommand{\XcOneTpchTenTpchKOneZeroSpeed}{1.25}

\newcommand{\XcOneTpchTenTpchKTwoFivePrimary}{1.09}

\newcommand{\XcOneTpchTenTpchKTwoFiveSpeed}{1.25}

\newcommand{\XeJobForestResidRand}{2.34}
\newcommand{\XeJobHTwoPoint}{9}

\newcommand{\XeJobKnnRand}{2.40}

\newcommand{\XeJobNnTplBelowFullKH}{14}
\newcommand{\XeJobNnTplBelowFullKOneZero}{19}
\newcommand{\XeJobNnTplOverFullKH}{0.98}
\newcommand{\XeJobNnTplOverFullKOneZero}{0.93}
\newcommand{\XeJobPoolMax}{4}
\newcommand{\XeJobPoolMin}{1}
\newcommand{\XeJobPooledRand}{2.40}

\newcommand{\XeJobPrimaryRand}{2.36}
\newcommand{\XeJobRandBelowRr}{18}
\newcommand{\XeJobTargetRand}{3.14}

\newcommand{\XeMoveMaxPct}{3}

\newcommand{\XeParallelTpch}{82}
\newcommand{\XeParallelTpchTen}{95}
\newcommand{\XeRoomJob}{1.05}
\newcommand{\XeRoomTpcds}{1.14}
\newcommand{\XeRoomTpch}{1.31}
\newcommand{\XeRoomTpchTen}{1.16}
\newcommand{\XeSpreadJob}{2.31}
\newcommand{\XeSpreadTpcds}{1.09}
\newcommand{\XeSpreadTpch}{1.30}

\newcommand{\XeSrcJobBoostingOracle}{2.00}
\newcommand{\XeSrcJobBoostingPrimary}{2.03}
\newcommand{\XeSrcJobBoostingSpeed}{2.11}

\newcommand{\XeSrcJobCostPrimary}{2.69}
\newcommand{\XeSrcJobCostSpeed}{4.11}
\newcommand{\XeSrcJobCostWins}{15}

\newcommand{\XeSrcJobOtherPairs}{80}
\newcommand{\XeSrcJobOtherWins}{19}

\newcommand{\XeSrcTpchTenOtherPairs}{24}
\newcommand{\XeSrcTpchTenOtherWins}{23}

\newcommand{\XeTpchTenHTwoPoint}{6}

\newcommand{\XeTpchTenNnTplOverFullKH}{1.00}

\newcommand{\XeTpchTenPrimaryKTwoFive}{1.16}

\newcommand{\XeTpchTenSpeedKTwoFive}{1.30}

\newcommand{\XAllHSixN}{26}

\newcommand{\XAllHSixOk}{0}

\newcommand{\XJobEpsilonTwoMax}{0.77}
\newcommand{\XJobEpsilonTwoMin}{0.28}

\newcommand{\XJobKH}{33}

\newcommand{\XJobKHControl}{2.40}
\newcommand{\XJobKHKnn}{3.03}
\newcommand{\XJobKHNnFull}{2.01}

\newcommand{\XJobKHNone}{2.71}
\newcommand{\XJobKHOracle}{2.36}
\newcommand{\XJobKHPrimary}{2.41}

\newcommand{\XJobKHSpeed}{2.48}

\newcommand{\XJobMOnetoMThreeShare}{0.49}

\newcommand{\XJobPairs}{20}

\newcommand{\XJobShareMax}{0.83}
\newcommand{\XJobShareMin}{0.49}
\newcommand{\XJobShareNoiseMax}{0.33}
\newcommand{\XJobShareNoiseMin}{0.28}

\newcommand{\XTpchTenHSixaN}{6}

\newcommand{\XTpchTenHSixaPoint}{3}

\newcommand{\XTpchTenHSixbMaxHi}{1.049}

\newcommand{\XTpchTenHThreeN}{6}
\newcommand{\XTpchTenHThreeOk}{2}

\newcommand{\XTpchTenHTwoVsSpeedMaxHi}{1.003}

\newcommand{\XTpchTenKHControl}{1.12}

\newcommand{\XTpchTenKHOracle}{1.11}
\newcommand{\XTpchTenKHPrimary}{1.12}
\newcommand{\XTpchTenKHRidge}{1.16}
\newcommand{\XTpchTenKHSpeed}{1.30}
\newcommand{\XTpchTenKOneZeroAffinePar}{1.31}

\newcommand{\XTpchTenKOneZeroSpeed}{1.31}

\newcommand{\XTpchTenPairs}{6}

\newcommand{\XTpchTenShareMax}{0.84}
\newcommand{\XTpchTenShareMin}{0.76}
\newcommand{\XTpchTenShareNoiseMax}{0.12}

\newcommand{\NNExtJobRrNnTplKThreeThree}{1.97}

\newcommand{\NNExtJobRrNnOracleKThreeThree}{1.99}

\begin{document}

\title{When Is One Query per Template Enough? Adapting Learned Cost Models to New Hardware}

\author{\IEEEauthorblockN{Krish Patel}
\IEEEauthorblockA{University of Waterloo, Waterloo, Canada\\
kmp.patel@uwaterloo.ca\quad ORCID 0009-0006-3064-0086}}

\maketitle

\begin{abstract}
A learned cost model that predicts query runtimes is trained on
measurements from one machine. When the database moves to other hardware, its plans stay the same
but their runtimes change, each by its own factor. We ask how to adapt such
a model from $k$ queries measured on the new machine: which queries to run,
how to use their runtimes, and how many are enough. On PostgreSQL we timed
the same plans of 1,162 TPC-H and TPC-DS queries in six environments: a
laptop and a server, each with all its processors or pinned to one, the
server also pinned to two, and the laptop also with a small buffer pool. No
single factor describes the slowdown between two environments: most of its
variance lies between query templates, so a speed factor stops improving
after a few queries. A correction that fits separate lines for serial and
parallel plans and adds an offset for every measured template brings the
moved model, with one measured query per template, to an error (geometric
mean over templates of the median Q-error) of
\PLadTpchLowdimTemplateRoundRobin{} on TPC-H and
\PLadTpcdsLowdimTemplateRoundRobin{} on TPC-DS, close to the same correction
on a model trained on the new machine; fine-tuning a DACE-style neural cost
model on the same queries gains at most \XtraNnGainPct\%. On templates never
measured, no correction we tried is on average more than \XtraHardBestPct\%
better than the speed factor. We then fixed hypotheses and timed new data.
On TPC-H at ten times the scale the errors follow these results, but two of
five repeated tests miss. On the Join Order Benchmark, whose queries differ
widely within a template, the model errs by \XJobKHOracle{} even when
retrained on the new machine, no correction gains much, and only the test
on new templates holds; boosted trees and a neural model fit these queries
better, and correcting the neural model works as well as fine-tuning it. We
report every test as registered.
\end{abstract}


\section{Introduction}
\label{sec:intro}
Predicting how long a query will run before it runs serves admission
control, scheduling and provisioning. Learned cost
models predict runtimes from plan features far better than raw optimizer
costs when they are trained on the system they predict
\cite{ganapathi2009,akdere2012,marcus2019,sun2020}. A runtime, however, is
a property of a plan on a machine, and machines change: a database moves to
new hardware, to a cloud instance of another size, or into a container with
fewer processors. The plans stay, and every runtime the model learned from
is now off, by an amount that depends on the query.

Systems handle a move in three ways. Some retrain on the new machine: T3
reruns its whole benchmark \cite{rieger2025t3}, QCFE reruns simplified
queries to refit its per-operator coefficients and then retrains on 200
labeled queries \cite{yan2024qcfe}, and DACE fine-tunes on a workload rerun
on the new machine \cite{liang2024dace}. Some map every old prediction
through one fitted line, the model shift used to move performance models of
configurable software \cite{valov2017,jamshidi2017}, whose simplest form
rescales every prediction by one speed factor. Others describe the machine
with features and train one model over many machines, as Redshift's Stage
predictor does with the instance type \cite{wu2024stage}. This paper asks
what lies between retraining and rescaling: if we can afford to run $k$
queries on the new machine, which queries should they be, how should their
runtimes be used, and how large must $k$ be?

We study the question with the plans held fixed, so that every change in
runtime comes from the environment. We timed 1,162 queries from 115
templates of TPC-H and TPC-DS on PostgreSQL in six environments that run
identical plans on copies of the same database. Their 30 ordered pairs per
workload change the processor count, the hardware, both, the buffer pool,
or several of these at once. To test whether the findings hold beyond
these data, we then fixed our hypotheses and measured new data: TPC-H at
scale factor 10 and the Join Order Benchmark (JOB) on its real IMDB data,
each in several environments. We contribute:
\begin{itemize}
\item A dataset of identical plans timed in many environments, which as far
  as we know has no public counterpart, and a measurement of how runtimes
  shift when only the environment changes (Section~\ref{sec:shift}). The
  slowdown differs widely between templates. Whether a plan has parallel
  workers explains much of a change of processor count, and how long a query
  runs explains much of a move from laptop to server; per-operator timings
  point to the mechanisms.
\item A comparison of six corrections, five baselines and a no-move
  control, a fine-tuned
  transformer cost model and six query selection strategies for 5 to 200
  measured queries (Sections~\ref{sec:adapt} to \ref{sec:neural}). On
  TPC-H and TPC-DS, one measured query per template with a per-template
  offset brings the moved model close to the same correction on a model
  trained on the new machine; refitting the model on the measured queries
  is far worse up to one query per template, also when the refit is told
  the templates, and fine-tuning a neural model on the same queries gains
  at most \XtraNnGainPct\%. In all four workloads of both rounds the moved
  model, so corrected, is on average within \XeMoveMaxPct\% of the same
  correction on a model trained on the new machine.
\item A negative result for new templates (Section~\ref{sec:unseen}),
  repeated on the new data: no correction we tried does much better than
  the speed factor, and its error is close to that of retraining. The
  remaining error is the model's error on a new template; the move adds
  little to it.
\item Two rounds of confirmatory tests (Sections~\ref{sec:ext}
  and~\ref{sec:confirm}). We committed hypotheses, the primary method and
  the neural baseline to our repository before the labels they are tested
  on existed, and report every test under its registered rule. The first
  round's five tests hold. The second round's mostly fail: at ten times the
  data the errors follow the first round, but two of the five repeated
  tests miss, and so does H6a; on the Join Order Benchmark, where a
  template's queries differ in runtime by a factor of \XeSpreadJob, the
  model's own error leaves little for any correction to gain, and only the
  test on new templates holds.
\item Recommendations for moving a runtime model (Section~\ref{sec:lessons})
  and a public artifact with every query, plan, label and script.
\end{itemize}

\section{Problem and Background}
\label{sec:problem}

\textbf{Setting.} A model $f_A$ predicts the log runtime of a query from
features $x$ of its PostgreSQL plan before the query runs, and was trained
on the runtimes of a pool of queries measured on machine $A$. The database
moves to machine $B$ with the same data, statistics and planner settings, so
the planner chooses the same plans; with runtimes $y_A$ and $y_B$ on $A$
and $B$, we call $y_B/y_A$ a query's slowdown.
Before $f_A$ is used on $B$ we may run $k$ queries there. Every query
belongs to a template, the parameterized SQL it was generated from; in a
production system templates can be read from query fingerprints such as
PostgreSQL's query identifier, as workload forecasters do
\cite{ma2018querybot}.

In machine-learning terms the inputs do not change: the same queries get
the same plans and features everywhere, so the distribution of $x$ is fixed
and only the distribution of runtime given $x$ shifts. Reweighting
unlabeled inputs, the usual answer to covariate shift
\cite{shimodaira2000}, has nothing to correct; the model needs runtimes
measured on $B$, and each costs time there.

\textbf{The shift.} On $B$, the source model's log error on a query $q$ is
\begin{equation}
e(q) = \log y_B(q) - f_A(x(q)) = \delta(q) + u_A(q),
\label{eq:error}
\end{equation}
the log slowdown $\delta(q) = \log(y_B(q)/y_A(q))$ plus the model's
residual $u_A(q)$ on $A$. A correction estimates $\delta$ from the $k$
measured log errors. The corrections below assume that $\delta$ is a constant
(a speed factor), a function of $x$, or a value per template. Which of
these fits a move is a property of the data that we can measure: the
template share
\begin{equation}
\eta^2 = \frac{\sum_{\tau} n_\tau(\bar\delta_\tau-\bar\delta)^2}{\sum_q(\delta(q)-\bar\delta)^2},
\label{eq:share}
\end{equation}
with $n_\tau$ queries of mean log slowdown $\bar\delta_\tau$ in template
$\tau$, is the part of the variance of $\delta$ that lies between
templates. Computed between two passes on one machine, it measures noise.
With few queries per template $\eta^2$ is biased upward, so we also report
$\omega^2=(SS_b-(T-1)\,MS_w)/(SS_t+MS_w)$, where $SS_b$ and $SS_t$ are the
numerator and denominator of \eqref{eq:share}, $T$ is the number of
templates and $MS_w$ the mean square within templates.

\textbf{Error measure.} We measure accuracy by the Q-error
$Q(y,\hat y)=\max(y/\hat y,\hat y/y)$ \cite{moerkotte2009}. A test
template's error is the median Q-error of its test queries. A method's
error is the geometric mean of these template errors, so that every
template counts once whatever its number of queries, averaged in log over
the random orders of the pool. A method beats another when the 95\% interval of the
geometric-mean ratio of their template errors, from resampling whole
templates \cite{efron1993}, lies below 1.

\section{Methods Compared}
\label{sec:methods}

\textbf{Features and source models.} The 19 plan features are the root's
startup and total cost, estimated rows and width; the number of nodes and
the depth; the largest row estimate of any node and the sum of the nodes'
log row estimates; the counts of all scans, sequential scans, index scans
(bitmap scans included), nested-loop, hash and merge joins, sorts,
aggregates and materializations; the workers planned over all Gather nodes;
and the number of subplans. Costs, rows and widths enter as $\log(1+x)$.
The primary source model, the \emph{forest model}, fits a line in $\log(1+\text{total cost})$ to the
log runtime (the cost line) and a random forest to its residuals (300
trees, at least two queries per leaf, 4 of the 19 features tried at each
split) \cite{breiman2001,pedregosa2011}, so that it falls back to the cost
line where the forest learned little. We also use a random forest alone,
gradient-boosted trees, ridge regression and the cost line alone, and a
neural plan model (below).

\textbf{Neural plan model.} Learned cost models in the literature read the
plan tree, not a fixed feature vector. As their representative we
reimplemented the architecture of DACE \cite{liang2024dace}, built to move
across databases and adapted to new hardware by fine-tuning. A plan is the
sequence of its nodes in depth-first order; each node is a one-hot node type
and the log optimizer estimates of its total and startup cost, rows and
width, with its height and planned workers, scaled by median and
interquartile range as in DACE. One post-norm transformer layer with one
head and width \NNDevHidden{} attends under a tree mask, so that a node sees
only its own subtree, and a multilayer perceptron maps the root to its log
runtime. DACE trains on every sub-plan's runtime; we have root runtimes only
and train on them, with the mean Q-error as loss. The model has
\NNDevTpchParams{} parameters on TPC-H and \NNDevTpcdsParams{} on TPC-DS.
Its learning rate (\NNDevLr), number of epochs (\NNDevEpochs) and dropout
(\NNDevDropout) were chosen by five-fold cross-validation over templates
inside the source pools of the development pairs (the 60 ordered pairs of
the SF1 data), and a fit takes about \NNDevSourceSeconds\,s
on two processor threads.

\textbf{Corrections.} All corrections act on the log errors of the $k$
measured queries, $e_i = \log y_i - f_A(x_i)$, and predict
$\log \hat y = f_A(x) + c(x)$ on $B$; fitting a learner to the residuals of
a base model follows \cite{didona2015}.
\begin{itemize}
\item \emph{Speed factor:} $c$ is the mean $\bar e$ of the $e_i$.
\item \emph{Ridge correction:} ridge regression \cite{hoerl1970} of $e$ on
  the 19 features standardized over the $k$ queries,
  \begin{equation}
  (\hat\beta_0,\hat\beta)=\arg\min_{\beta_0,\beta}\ \textstyle\sum_{i=1}^k\bigl(e_i-\beta_0-\beta^\top\tilde x_i\bigr)^2+\lambda\|\beta\|^2,
  \label{eq:ridge}
  \end{equation}
  with an unpenalized intercept and $\lambda$ chosen by leave-one-out
  \cite{golub1979} from 45 values between $10^{-6}$ and $10^{5}$.
\item \emph{Affine+parallel:} $c(x) = a + b\,f_A(x) + \pi(x)\,(a' + b'\,f_A(x))$,
  where $\pi(x)$ is 1 for plans with parallel workers; $b$ makes the
  slowdown depend on the predicted runtime. The intercept $a'$ enters only
  when both kinds of plan are among the $k$ queries, the slope $b'$ only
  when each kind has two. It is fitted as in \eqref{eq:ridge}, with
  $f_A(x)$, $\pi(x)$ and $\pi(x)f_A(x)$ as features.
\item \emph{Template offsets:} the speed factor plus, for every template
  among the $k$ queries, the mean residual of its queries from that factor;
  other templates get the factor alone.
\item \emph{Template correction}, our primary method: affine+parallel plus,
  for every template among the $k$ queries, the mean residual of its
  queries from that line; other templates get the line alone.
\item \emph{1-NN:} the log error of the nearest of the $k$ queries in the
  standardized feature space. It needs no template labels.
\end{itemize}
The speed factor is one end of the ridge correction's family:
\begin{proposition}\label{prop:limit}
For every $\lambda>0$, the solution of \eqref{eq:ridge} has
$\hat\beta_0=\bar e$ and
$\hat\beta=(\tilde X^\top\tilde X+\lambda I)^{-1}\tilde X^\top(e-\bar e\mathbf 1)$,
where $\tilde X$ stacks the $\tilde x_i$, and
$\|\hat\beta\|\le\|\tilde X^\top(e-\bar e\mathbf 1)\|/\lambda$. As
$\lambda\to\infty$ the ridge correction becomes the speed factor.
\end{proposition}
\begin{proof}
The standardized features sum to zero over the $k$ queries, so the
derivative in $\beta_0$ gives $k\beta_0=\sum_i e_i=k\bar e$. Then $\beta$
minimizes $\|(e-\bar e\mathbf 1)-\tilde X\beta\|^2+\lambda\|\beta\|^2$,
with the stated minimizer, and the eigenvalues of
$(\tilde X^\top\tilde X+\lambda I)^{-1}$ are at most $1/\lambda$.
\end{proof}
Leave-one-out chooses how far to move from the speed factor toward a full
fit of the $k$ log errors; with few queries this choice is noisy. Template offsets
put all their freedom in the templates instead.

\textbf{Baselines.} \emph{Target only} refits the forest model on the $k$
queries alone; the \emph{pooled refit} refits it on the source pool and the
$k$ queries with the machine as a feature; the \emph{forest on the errors}
fits a forest to the $e_i$, and, to check whether a refit that knows the
templates does better, also to the $e_i$ with one indicator per template; and a \emph{template lookup} predicts a query by
the mean measured log runtime of its template, with no model. For the
neural model, \emph{fine-tuning} trains all its weights on the $k$ queries,
and \emph{head fine-tuning} only the perceptron, the part DACE adapts when
it moves to new hardware; their learning rate and number of steps depend on
$k$ only and were chosen on the development pairs, scored on target pool
queries outside the $k$. The \emph{oracle} refits the forest model on
the whole target pool. It is a reference, not a bound: it knows nothing of
templates. The \emph{no-move control} applies a method to a model trained
on the target pool itself, with the same $k$ queries; the gap to it is the
cost of the move.

\textbf{Choosing the $k$ queries.} \emph{Random} takes them in a random
order. \emph{Round robin} takes one query of each template in turn, in a
random order of templates and of queries; \emph{cheapest per template} is
round robin with the query that ran fastest on $A$ first. \emph{k-center}
adds the query farthest from those chosen so far in the standardized
feature space \cite{gonzalez1985,sener2018}. \emph{D-optimal} adds the query
that most increases $\det(I + X^\top X)$, where $X$ holds the chosen
queries' standardized features and a column of ones
\cite{wynn1970,mitchell1974}. \emph{Plan class} alternates between parallel
and serial plans. Every strategy derives its $r$-th sequence from the $r$-th
of 20 random orders, so comparisons are paired; none looks at a runtime on
$B$.

\section{Experimental Methodology}
\label{sec:setup}

\textbf{Environments.} M1 is a laptop (Intel Core Ultra 7 155H, 22
threads) running PostgreSQL 18.6 in WSL2 (a Linux virtual machine on
Windows) with 16\,GB of memory; M3 is a virtual machine with 64 virtual
processors of an AMD EPYC 9754 and 1\,TB of memory on a shared university
server. M2 and M4 are M1 and M3 with every PostgreSQL server process pinned
to one virtual processor, so that a plan's leader and its parallel workers
(at most two per Gather node) share it while the client keeps all
processors; M6 is M3 pinned to two. M5 is M1 with PostgreSQL's default
buffer pool, 128\,MB instead of 2\,GB. All use the same PostgreSQL build,
which has no JIT, and the same settings (work\_mem 64\,MB,
effective\_cache\_size 8\,GB) apart from M5's buffer pool, on copies of the
same database files; every saved plan, with its estimates and costs, is
identical in all environments of a workload. The data at scale factor 1
(SF1) fit in memory everywhere. Table~\ref{tab:env} lists the environments.

\begin{table}[t]
\caption{The environments. CPUs: the processors that PostgreSQL's server processes may use (the client keeps all). Buffers: PostgreSQL's buffer pool. Noise: the median Q-error between two passes of the same query; an empty cell is a workload the environment did not run.}
\label{tab:env}
\small
\setlength{\tabcolsep}{4pt}
\centering
\begin{tabular}{@{}llllrrrr@{}}
\toprule
& & & & \multicolumn{4}{c}{noise, median Q} \\
\cmidrule(l){5-8}
Env. & machine & CPUs & buffers & TPC-H & TPC-DS & SF10 & JOB \\
\midrule
M1 & laptop & all 22 & 2\,GB & 1.07 & 1.09 &  & 1.06 \\
M2 & laptop & 1 & 2\,GB & 1.07 & 1.08 &  & 1.03 \\
M3 & server & all 64 & 2\,GB & 1.05 & 1.05 & 1.05 & 1.05 \\
M4 & server & 1 & 2\,GB & 1.07 & 1.03 & 1.07 & 1.04 \\
M5 & laptop & all 22 & 128\,MB & 1.03 & 1.05 &  &  \\
M6 & server & 2 & 2\,GB & 1.08 & 1.04 & 1.05 & 1.05 \\
\bottomrule
\end{tabular}

\end{table}

\textbf{Workloads.} TPC-H at scale factor 1 \cite{tpch}: 680 queries from
22 templates, 420 from its query generator and 260 with range parameters
drawn wider than the specification's, so that runtimes vary inside a
template. TPC-DS at scale factor 1 \cite{tpcds}: 482 queries from the 93
templates whose first query ran within 30\,s on M1; in its four templates of
two statements, each statement counts as a query of the template. Queries
are split once, inside every template, into a pool from which the $k$
queries are drawn (476 and 294 queries, 70\% and 61\%) and a test set on
which the corrections are scored. We link two TPC-DS templates when their
first queries, with literals masked, share at least 85\% of their tokens,
and take the connected components as families: 75 families, 15 of them with
more than one template; intervals on TPC-DS resample families, so that near
copies do not count as independent.

The extension adds two workloads, each on copies of one cluster. TPC-H at
scale factor 10 (17\,GB with indexes) uses 220 of the SF1 queries, 7 pool
and 3 test queries per template drawn with a fixed seed from the SF1 split,
with Q11's fraction divided by the scale factor as the specification
prescribes; it runs on the server as M3, M6 and M4.
JOB \cite{leis2015} runs its 113 queries from 33 templates on the IMDB
snapshot of its authors (8.7\,GB with its foreign-key indexes), split inside
every template (78 pool and 35 test queries), on the laptop (M1, M2) and the
server (M3, M6, M4). Every JOB query ran once on the server before any
label, and all finished within the 60\,s limit we had fixed.

\textbf{Labels.} Every environment runs every query in three passes in
different random orders; in a pass a query runs four times with
\texttt{EXPLAIN (ANALYZE, TIMING OFF)}, the first only to warm the cache. A
run's time is the execution time PostgreSQL reports, without planning or
sending rows to the client, and a label is the median over passes of the
median of the three timed runs. A pass whose level shifts by more than 10\%
partway through, or during which the laptop slept or ran on battery, is
measured once more. We set the clock of the laptop's virtual machine to
track the host's, after finding that it ran fast, and record its rate in
every pass; the laptop ran nothing else during a pass, and the shared
server's passes started at night once its load was low. The median Q-error between two passes of the same query
is \EnvNoiseMedQMin{} to \EnvNoiseMedQMax{} per environment and workload
(Table~\ref{tab:env}). For
Section~\ref{sec:shift} we also ran every SF1 query once more in M1 to M5
with per-node timing on, after a warm-up; an operator's own time is its time
over all loops minus its children's, and these timings are never labels.

\textbf{Protocol.} For every ordered pair, the source model is trained on
$A$'s pool, the $k$ queries are drawn from $B$'s pool for $k$ from 5 to 200
(in the second round to 100 on SF10 and 50 on JOB, whose pools are smaller)
over 20 random orders, and every correction is scored on $B$'s test set.
The $k$ queries are thus queries the source model was trained on, as in
practice, where the queries rerun on a new machine come from the workload
the model learned from; the no-move control is in the same position. The
choice does not drive the comparison: with the source model trained on half
of each template's pool queries and the $k$ queries (one per template)
drawn from the other half instead of the same half, the primary method errs
by \XhoTpchOutPrimary{} instead of \XhoTpchInPrimary{} on TPC-H and
\XhoTpcdsOutPrimary{} instead of \XhoTpcdsInPrimary{} on TPC-DS, and the
no-move control moves alike (\XhoTpchOutControl{} instead of
\XhoTpchInControl, \XhoTpcdsOutControl{} instead of \XhoTpcdsInControl). The
test on new templates holds out one template at a time (in TPC-DS its
family): the source model is retrained on $A$'s pool without it, the $k=25$
queries are drawn from $B$'s pool without it, the oracle is retrained on
that pool, and the template's test queries are scored. Queries whose plan
features differ between the two environments of a pair would be left out;
none did.

\textbf{Exploratory and confirmatory analyses.} We designed the corrections
and strategies on TPC-H with M1, M2 and M3, so those pairs are exploratory.
We committed the primary method (the template correction with round-robin
selection) and five hypotheses (commit 17c328f) after the first TPC-DS
passes on M2 and M3 had started, but before any label of TPC-DS on M2, M3
or M4 or of TPC-H on M4 existed. Within twenty minutes, still before any
such label, a code review led to an amendment (commits b2e59dc and
fc9dc9a): TPC-DS intervals resample families, the test on new templates
holds out families, and affine+parallel and the primary method leave out
their parallel terms when the $k$ queries hold one plan class. For the
extension we committed the new workloads, the same five hypotheses for
their pairs and a sixth on the neural baseline (commit 13f14ce) before any
of their queries ran. We then froze the neural model and its fine-tuning
rules on the development pairs (commit 52cc5b4), and amended the sixth
hypothesis after its development results, before any new pair was
evaluated (commit 909c069). Everything else is exploratory: the
per-operator timings, M5 and, in the first round, M6, the no-move control, the lookup, the
single-run labels, the other source models, streams, shift transfer,
three analyses we added after an internal review (the forest on the errors
with template indicators, the split of the pool into halves, and the
tails), and the numbers that explain the second round's results, which we
computed after seeing them (Section~\ref{sec:ext}; the Lessons and the
Conclusion use them too).

\textbf{Artifacts.} The queries, plans, labels, per-operator timings, the
study plan with its timestamped amendments, and the scripts that build
every number, table and figure of this paper are public at
\url{https://github.com/KrishPatel13/query-runtime-transfer}. A fresh clone
reproduces every reported number of the SF1 study exactly on Linux.

\section{Results}
\label{sec:results}

\subsection{How runtimes shift}
\label{sec:shift}

\begin{table*}[t]
\caption{How the runtime of the same plan changes, for seven of the 30 ordered pairs of each workload. Slowdown: median of target over source runtime, for serial plans and plans with parallel workers. Template share: the share of the variance of the log slowdown between templates (shift), the same corrected for the number of templates ($\omega^2$), and the share between two passes on the source alone (noise). $R^2$: least-squares fits of the log slowdown on whether the plan is parallel, on the log source runtime (size) and on both with their interaction; features (new): a ridge model on the 19 features scored on templates (TPC-DS: families) left out of its fit, which can be negative.}
\label{tab:shift}
\small
\begin{tabular}{@{}llrrrrrrrrr@{}}
\toprule
& & \multicolumn{2}{c}{Slowdown} & \multicolumn{3}{c}{Template share} & \multicolumn{4}{c}{$R^2$ of the log slowdown on} \\
\cmidrule(lr){3-4}\cmidrule(lr){5-7}\cmidrule(l){8-11}
Pair & Change & serial & parallel & shift & $\omega^2$ & noise & parallel & size & both & features (new) \\
\midrule
\multicolumn{11}{@{}l}{\emph{TPC-H}} \\
M1$\to$M2 & processors & 1.12 & 2.66 & 0.86 & 0.86 & 0.08 & 0.61 & 0.00 & 0.61 & 0.27 \\
M3$\to$M4 & processors & 1.01 & 2.07 & 0.84 & 0.83 & 0.03 & 0.57 & 0.12 & 0.62 & 0.53 \\
M3$\to$M6 & two processors & 1.00 & 1.32 & 0.69 & 0.68 & 0.03 & 0.51 & 0.06 & 0.55 & 0.31 \\
M1$\to$M3 & hardware & 1.81 & 2.07 & 0.80 & 0.79 & 0.08 & 0.07 & 0.52 & 0.63 & 0.45 \\
M2$\to$M4 & hardware & 1.63 & 1.69 & 0.74 & 0.73 & 0.03 & 0.02 & 0.22 & 0.33 & 0.41 \\
M1$\to$M4 & both & 1.82 & 4.42 & 0.91 & 0.90 & 0.08 & 0.54 & 0.03 & 0.60 & 0.26 \\
M1$\to$M5 & buffer pool & 1.47 & 1.07 & 0.72 & 0.71 & 0.08 & 0.03 & 0.03 & 0.08 & -0.12 \\
\multicolumn{11}{@{}l}{\emph{TPC-DS}} \\
M1$\to$M2 & processors & 1.06 & 2.11 & 0.91 & 0.89 & 0.18 & 0.44 & 0.00 & 0.50 & 0.26 \\
M3$\to$M4 & processors & 0.92 & 2.05 & 0.94 & 0.93 & 0.18 & 0.56 & 0.10 & 0.59 & 0.38 \\
M3$\to$M6 & two processors & 0.89 & 1.25 & 0.88 & 0.85 & 0.18 & 0.33 & 0.19 & 0.44 & 0.30 \\
M1$\to$M3 & hardware & 2.06 & 1.80 & 0.83 & 0.80 & 0.18 & 0.07 & 0.64 & 0.65 & 0.20 \\
M2$\to$M4 & hardware & 1.80 & 1.77 & 0.85 & 0.81 & 0.22 & 0.01 & 0.41 & 0.45 & -0.05 \\
M1$\to$M4 & both & 1.83 & 3.79 & 0.92 & 0.90 & 0.18 & 0.34 & 0.06 & 0.52 & 0.19 \\
M1$\to$M5 & buffer pool & 0.98 & 1.03 & 0.77 & 0.72 & 0.18 & 0.00 & 0.02 & 0.02 & -0.06 \\
\bottomrule
\end{tabular}

\end{table*}

Table~\ref{tab:shift} shows seven pairs per workload. No single factor
describes any of them. Over all pairs, a share of \PSRTpchShareMin{} to
\PSRTpcdsShareMax{} of the variance of the log slowdown lies between
templates (\PSRTpchEpsilonTwoMin{} to \PSRTpcdsEpsilonTwoMax{} after
correcting for the number of templates), against at most \PSRTpchNoiseMax{}
(TPC-H) and \PSRTpcdsNoiseMax{} (TPC-DS) between two passes of one machine.
What separates the templates depends on the change. Pinning PostgreSQL to
one processor gives plans with parallel workers a median slowdown of
\PSTpcdsMThreetoMFourParallelX{} to \PSTpchMOnetoMTwoParallelX{} and serial
plans one of \PSTpcdsMThreetoMFourSerial{} to \PSTpchMOnetoMTwoSerial, and
that one bit explains
\PSTpcdsMOnetoMTwoRTwoParallel{} to \PSTpchMOnetoMTwoRTwoParallel{} of the
variance; with two processors (M3$\to$M6) parallel plans slow by only
\PSTpchMThreetoMSixParallelX{} (TPC-H) and \PSTpcdsMThreetoMSixParallelX{}
(TPC-DS). Moving from the laptop to the server slows serial and parallel
plans alike, but short queries more than long ones: the log slowdown falls
as the log runtime grows, and size explains \PSTpchMOnetoMThreeRTwoSize{}
(TPC-H) and \PSTpcdsMOnetoMThreeRTwoSize{} (TPC-DS) of its variance from M1
to M3 and \PSTpchMTwotoMFourRTwoSize{} and \PSTpcdsMTwotoMFourRTwoSize{}
from M2 to M4. The small buffer pool gives a median slowdown of only
\PSTpchMOnetoMFiveSlowdown{} (TPC-H) and \PSTpcdsMOnetoMFiveSlowdown{}
(TPC-DS), although TPC-H's serial plans slow by \PSTpchMOnetoMFiveSerial,
and parallelism and size together explain only \PSTpchMOnetoMFiveRTwoBoth{}
and \PSTpcdsMOnetoMFiveRTwoBoth{} of its variance.

\begin{table}[t]
\caption{Median ratio of each operator's own time (target over source), inside (par.) and outside (ser.) the parallel part of plans, over the nodes of both workloads weighted by the source's time, from one extra pass with per-node timing. Proc.: one processor (M1$\to$M2, M3$\to$M4); hardware: M1$\to$M3; buffer pool: M1$\to$M5.}
\label{tab:ops}
\footnotesize
\setlength{\tabcolsep}{2.4pt}
\begin{tabular}{@{}lrrrrrrrr@{}}
\toprule
& \multicolumn{2}{c}{Proc. (laptop)} & \multicolumn{2}{c}{Proc. (server)} & \multicolumn{2}{c}{Hardware} & \multicolumn{2}{c}{Buffer pool} \\
\cmidrule(lr){2-3}\cmidrule(lr){4-5}\cmidrule(lr){6-7}\cmidrule(lr){8-9}
Operator & par. & ser. & par. & ser. & par. & ser. & par. & ser. \\
\midrule
Seq Scan & 2.85 & 0.99 & 2.85 & 1.01 & 1.75 & 1.70 & 1.12 & 1.23 \\
Index Scan & 2.33 & 1.00 & 2.32 & 1.07 & 1.64 & 1.01 & 1.92 & 1.42 \\
Bitmap Heap Scan & 1.98 & 0.99 & 2.04 & 1.20 & 2.01 & 1.49 & 0.98 & 2.14 \\
Hash Join & 2.56 & 1.01 & 2.57 & 0.94 & 1.62 & 1.65 & 1.00 & 1.00 \\
Hash & 2.89 & 1.05 & 2.32 & 1.02 & 1.32 & 1.34 & 1.13 & 1.06 \\
Nested Loop & 1.50 & 1.01 & 0.93 & 0.91 & 1.56 & 1.72 & 1.09 & 1.03 \\
Sort & 2.53 & 1.01 & 2.59 & 1.02 & 1.25 & 1.43 & 1.04 & 1.07 \\
Hash aggregate & 2.94 & 1.12 & 3.05 & 1.15 & 1.43 & 1.50 & 0.98 & 1.08 \\
Sorted aggregate & 1.78 & 1.00 & 1.99 & 1.15 & 1.42 & 1.58 & 1.02 & 1.04 \\
\bottomrule
\end{tabular}

\end{table}

Per-operator timings point to the mechanisms (Table~\ref{tab:ops}),
although per-node timing adds overhead that differs between machines. On
one processor, the main operators of each workload (at least 3\% of its time
on the source; these ranges count operators per workload, while
Table~\ref{tab:ops} pools both) take \OpProcParallel{} times as long inside the parallel part
of a plan on the laptop and \OpServerProcParallel{} times on the server,
while serial operators keep their speed: the same mechanism on both
machines. The move from laptop to server slows operators unevenly, by
\OpHwRange{} times depending on the operator, so plans with different
operators slow differently. With the small buffer pool, index scans, which
revisit pages that no longer stay in PostgreSQL's buffers, slow down
\OpMemIndex{} times, while sequential scans, which read pages in order,
change by \OpMemSeq. Which templates slow down thus depends on their
operators, which plan-level features see only in part: a ridge model on the
19 features, fitted without the template it predicts, explains at most
\PSRTpchRTwoLotMax{} of the variance, and nothing for the buffer pool (last
column of Table~\ref{tab:shift}).

On the extension's data the template share is \XTpchTenShareMin{} to
\XTpchTenShareMax{} on SF10 (noise at most \XTpchTenShareNoiseMax) but
\XJobShareMin{} to \XJobShareMax{} on JOB ($\omega^2$ \XJobEpsilonTwoMin{}
to \XJobEpsilonTwoMax), whose few queries per template give a noise share
of \XJobShareNoiseMin{} to \XJobShareNoiseMax; on JOB the move between M1
and M3 has the lowest share (\XJobMOnetoMThreeShare). Moving to
fewer processors on the same machine again slows parallel plans, by
\XcNewTpchTenShrinkParallelMin{} to \XcNewTpchTenShrinkParallelMax{} on SF10
and \XcNewJobShrinkParallelMin{} to \XcNewJobShrinkParallelMax{} on JOB,
while serial plans change by \XcNewTpchTenShrinkSerialMin{} to
\XcNewTpchTenShrinkSerialMax{} and \XcNewJobShrinkSerialMin{} to
\XcNewJobShrinkSerialMax{} (medians per pair).

\begin{figure*}[t]
\centering
\includegraphics[width=\textwidth]{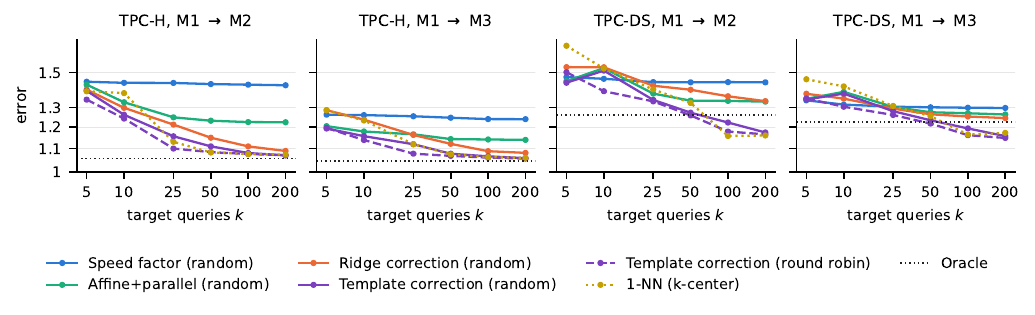}
\caption{Error (geometric mean over test templates of the median Q-error) against the number $k$ of measured target queries, both axes logarithmic, for a processor change (M1$\to$M2) and a hardware change (M1$\to$M3) on each workload; the TPC-H panels are pairs the methods were designed on. Line style gives the selection: solid random, dashed round robin, dotted k-center. The horizontal dotted line is the oracle.}
\label{fig:learning}
\end{figure*}

\begin{table*}[t]
\caption{Error by kind of change, as the geometric mean over its ordered pairs. Speed, Aff.+par.\ and Ridge use random selection; Primary (the template correction) uses random selection at $k=10$ and round robin at $k=25$ and $k=100$. Oracle: the forest model refitted on the whole target pool. New templates: each template (TPC-DS: its family) held out from the source model, the $k=25$ randomly chosen queries, the oracle and every correction. Mixed: the remaining 14 pairs, all involving M5 or M6.}
\label{tab:adapt}
\small
\setlength{\tabcolsep}{4.6pt}
\begin{tabular}{@{}lrrrrrrrrrrrr@{}}
\toprule
& & \multicolumn{3}{c}{$k=10$, random} & \multicolumn{2}{c}{$k=25$} & $k=100$ & & \multicolumn{4}{c}{New templates, $k=25$} \\
\cmidrule(lr){3-5}\cmidrule(lr){6-7}\cmidrule(lr){8-8}\cmidrule(l){10-13}
Change & Pairs & Speed & Aff.+par. & Primary & Ridge & Primary & Primary & Oracle & Speed & Aff.+par. & Ridge & Oracle \\
\midrule
\multicolumn{13}{@{}l}{\emph{TPC-H}} \\
processors & 4 & 1.41 & 1.28 & 1.23 & 1.20 & 1.09 & 1.07 & 1.05 & 1.99 & 2.00 & 2.13 & 1.98 \\
two processors & 2 & 1.14 & 1.12 & 1.11 & 1.12 & 1.08 & 1.06 & 1.05 & 1.76 & 1.76 & 1.77 & 1.76 \\
hardware & 4 & 1.22 & 1.19 & 1.17 & 1.15 & 1.09 & 1.07 & 1.05 & 1.99 & 1.98 & 1.99 & 1.98 \\
both & 4 & 1.46 & 1.33 & 1.26 & 1.21 & 1.10 & 1.08 & 1.05 & 2.01 & 2.01 & 2.11 & 1.98 \\
buffer pool & 2 & 1.38 & 1.39 & 1.32 & 1.33 & 1.13 & 1.10 & 1.05 & 2.16 & 2.14 & 2.47 & 2.11 \\
mixed & 14 & 1.44 & 1.35 & 1.29 & 1.29 & 1.11 & 1.08 & 1.05 & 2.08 & 2.03 & 2.27 & 2.03 \\
\multicolumn{13}{@{}l}{\emph{TPC-DS}} \\
processors & 4 & 1.52 & 1.49 & 1.47 & 1.49 & 1.35 & 1.18 & 1.25 & 2.72 & 2.65 & 2.73 & 2.63 \\
two processors & 2 & 1.36 & 1.34 & 1.33 & 1.33 & 1.28 & 1.16 & 1.23 & 2.55 & 2.54 & 2.59 & 2.52 \\
hardware & 4 & 1.34 & 1.35 & 1.35 & 1.34 & 1.28 & 1.18 & 1.25 & 2.64 & 2.64 & 2.67 & 2.63 \\
both & 4 & 1.55 & 1.51 & 1.49 & 1.55 & 1.37 & 1.18 & 1.25 & 2.72 & 2.65 & 2.74 & 2.63 \\
buffer pool & 2 & 1.47 & 1.49 & 1.48 & 1.46 & 1.41 & 1.24 & 1.31 & 3.01 & 3.01 & 3.05 & 2.99 \\
mixed & 14 & 1.50 & 1.49 & 1.47 & 1.50 & 1.38 & 1.19 & 1.27 & 2.78 & 2.75 & 2.85 & 2.73 \\
\bottomrule
\end{tabular}

\end{table*}

\subsection{Adapting with \texorpdfstring{$k$}{k} queries}
\label{sec:adapt}

\begin{table}[t]
\caption{Error at the budget where each template can have one query, as the geometric mean over the 30 ordered pairs of a workload; the last row applies the primary method to a model trained on the target itself (no move).}
\label{tab:ladder}
\small
\setlength{\tabcolsep}{3pt}
\begin{tabular}{@{}llrr@{}}
\toprule
Method & Selection & TPC-H & TPC-DS \\
& & $k=25$ & $k=100$ \\
\midrule
No adaptation &  & 1.83 & 1.96 \\
Speed factor & random & 1.37 & 1.45 \\
Ridge correction & random & 1.24 & 1.38 \\
Target only & random & 1.51 & 1.80 \\
Pooled refit & random & 1.36 & 1.36 \\
Forest on the errors & random & 1.22 & 1.28 \\
Template lookup (no model) & round robin & 1.24 & 1.24 \\
Template offsets & round robin & 1.12 & 1.19 \\
1-NN & round robin & 1.10 & 1.19 \\
Primary & round robin & 1.10 & 1.19 \\
Oracle (whole target pool) &  & 1.05 & 1.26 \\
\midrule
No move: primary & round robin & 1.08 & 1.18 \\
\bottomrule
\end{tabular}

\end{table}

Figure~\ref{fig:learning} and Table~\ref{tab:adapt} give the errors, and
Table~\ref{tab:ladder} compares the methods at one query per template.
Without adaptation the source model errs by \PNoneTpchKTwoFive{} on TPC-H
and \PNoneTpcdsKOneZeroZero{} on TPC-DS. The speed factor finds the average
slowdown from a few queries and then stays flat (TPC-H: \PSpeedTpchKFive{}
at $k=5$, \PSpeedTpchKTwoFive{} at $k=25$; TPC-DS: \PSpeedTpcdsKFive{} and
\PSpeedTpcdsKTwoFive), because no single factor fits templates whose
slowdowns differ. Affine+parallel helps at small budgets on TPC-H: it beats
the speed factor at $k=10$ in \PAffTPCHBelow{} of \PAffTPCHPairs{} pairs and
loses in \PAffTPCHAbove; for processor changes it lowers the $k=10$ error
from \PATpchProcessorsRandomSpeedOneZero{} to
\PATpchProcessorsRandomAffineParallelOneZero, and the primary method's
template offsets bring it to \PATpchProcessorsRandomLowdimTemplateOneZero.
On TPC-DS, affine+parallel beats the speed factor at $k=10$ in
\PAffTPCDSBelow{} of \PAffTPCDSPairs{} pairs and loses in \PAffTPCDSAbove, and its lead grows
with $k$ (Figure~\ref{fig:learning}).

Once each template has a measured query, the template offsets carry the
gain. With round-robin selection the primary method reaches
\PLadTpchLowdimTemplateRoundRobin{} on TPC-H at $k=25$ and
\PLadTpcdsLowdimTemplateRoundRobin{} on TPC-DS at $k=100$; offsets on the
speed factor alone give \PLadTpchTemplateRawRoundRobin{} and
\PLadTpcdsTemplateRawRoundRobin, so at this budget the affine+parallel
line adds little. The same
correction on a model trained on the target itself reaches
\PLadTpchLowdimTemplateControlRoundRobin{} and
\PLadTpcdsLowdimTemplateControlRoundRobin: the move costs little. Without
the correction, that model is the oracle, which refits the template-blind
forest model on the whole target pool. On TPC-H the oracle errs by
\PLadTpchOracleRandom, a little less than the no-move control, so there the
correction's own noise costs a little. On TPC-DS the oracle errs by
\PLadTpcdsOracleRandom, since three pool queries per template are too few
for its forest, and the primary method's error is \PPrimOracleTPCDSMin{} to
\PPrimOracleTPCDSMax{} times the oracle's in every pair. The model matters inside a template: a lookup of the measured
runtime per template errs by \PLadTpchLookupRoundRobin{} on TPC-H against
\PLadTpchLowdimTemplateRoundRobin{} for the primary method, because the
widened queries vary inside their templates; on TPC-DS, whose queries vary
inside a template by a factor of only \PSpreadTPCDS{} (median geometric
standard deviation), the gap is smaller (\PLadTpcdsLookupRoundRobin{}
against \PLadTpcdsLowdimTemplateRoundRobin). The refits and the forest on
the errors ran with random selection only. With the same selection the
primary method reaches \PLTpchKTwoFiveRandomLowdimTemplate{} and
\PLTpcdsKOneZeroZeroRandomLowdimTemplate{} (Table~\ref{tab:select}), against
\PLadTpchTargetRandom{} and \PLadTpcdsTargetRandom{} for target only,
\PLadTpchPooledRandom{} and \PLadTpcdsPooledRandom{} for the pooled refit,
and \PLadTpchForestResidRandom{} and \PLadTpcdsForestResidRandom{} for a
forest on the errors. Knowing the templates does not rescue the refit: with
one indicator per template the forest on the errors errs by
\XtraTpchForestTplRrKTwoFive{} and \XtraTpcdsForestTplRrKOneZeroZero{}
under round robin, against \XtraTpchPrimaryRrKTwoFive{} and
\XtraTpcdsPrimaryRrKOneZeroZero{} for the primary method, which it beats in
\XtraTpchForestTplWinsRrKTwoFive{} and \XtraTpcdsForestTplWinsRrKOneZeroZero{}
of 30 pairs, since with about one query per template a forest has little
to split each template's indicator on. The refits catch up only with far more queries than
templates: at $k=200$ on TPC-H, target only errs by
\XtraTpchTargetRandKTwoZeroZero{} and the forest on the errors by
\XtraTpchForestResidRandKTwoZeroZero, against
\XtraTpchPrimaryRandKTwoZeroZero{} for the primary method (all random).

The tails follow the medians. At one query per template, the 90th
percentile of the per-query Q-error over the test queries is
\XtlTpchPNinetyPrimary{} for the primary method against
\XtlTpchPNinetySpeed{} for the speed factor on TPC-H (oracle
\XtlTpchPNinetyOracle) and \XtlTpcdsPNinetyPrimary{} against
\XtlTpcdsPNinetySpeed{} on TPC-DS (oracle \XtlTpcdsPNinetyOracle). The
primary method's median prediction is off by a factor of only
\XtlTpchBiasPrimary{} (TPC-H) and \XtlTpcdsBiasPrimary{} (TPC-DS), against
\XtlTpchBiasSpeed{} and \XtlTpcdsBiasSpeed{} for the speed factor. On
TPC-DS a few test queries stay far off for every method: the 99th
percentile is \XtlTpcdsPNinetyNinePrimary{} for the primary method and
\XtlTpcdsPNinetyNineOracle{} for the oracle.

\subsection{Which queries to measure}
\label{sec:select}

\begin{table}[t]
\caption{Error of three corrections (columns) when the $k$ queries are chosen by each strategy (rows), as the geometric mean over the 30 ordered pairs of a workload, at the smallest budget of our grid that holds one query per template. Seconds: total runtime of the $k$ queries on the target, one run each.}
\label{tab:select}
\small
\begin{tabular}{@{}lrrrr@{}}
\toprule
Strategy & Primary & 1-NN & Ridge & Seconds \\
\midrule
\multicolumn{5}{@{}l}{\emph{TPC-H, $k=25$}} \\
random & 1.17 & 1.19 & 1.24 & 22 \\
round robin & 1.10 & 1.10 & 1.17 & 24 \\
cheapest per template & 1.17 & 1.17 & 1.23 & 16 \\
k-center & 1.13 & 1.13 & 1.21 & 21 \\
D-optimal & 1.14 & 1.13 & 1.20 & 16 \\
plan class & 1.19 & 1.21 & 1.26 & 18 \\
\multicolumn{5}{@{}l}{\emph{TPC-DS, $k=100$}} \\
random & 1.23 & 1.29 & 1.38 & 127 \\
round robin & 1.19 & 1.19 & 1.37 & 110 \\
cheapest per template & 1.22 & 1.23 & 1.38 & 88 \\
k-center & 1.18 & 1.18 & 1.36 & 132 \\
D-optimal & 1.28 & 1.39 & 1.35 & 188 \\
plan class & 1.24 & 1.30 & 1.41 & 118 \\
\bottomrule
\end{tabular}

\end{table}

Table~\ref{tab:select} compares the strategies. Covering the templates is
what matters. On TPC-H at $k=25$, round robin lowers the primary method's
error from \PLTpchKTwoFiveRandomLowdimTemplate{} to
\PLTpchKTwoFiveRoundRobinLowdimTemplate. Without template labels, k-center
selection with the 1-NN correction comes close on TPC-H
(\PLTpchKTwoFiveKCenterOneNn) and matches the primary method on TPC-DS at
$k=100$ (\PLTpcdsKOneZeroZeroKCenterOneNn{} against
\PLTpcdsKOneZeroZeroRoundRobinLowdimTemplate, lower in \PNnBelowPrimTPCDS{}
of 30 pairs), since queries of one template have similar plans. D-optimal selection,
which is built for the ridge correction, gives that correction its lowest
TPC-DS error, but on TPC-DS it picks long queries
(\PLTpcdsKOneZeroZeroDOptimalCalibrationS\,s against
\PLTpcdsKOneZeroZeroRoundRobinCalibrationS\,s for round robin) and worsens
the primary method; on TPC-H it picks the cheapest set
(\PLTpchKTwoFiveDOptimalCalibrationS\,s against
\PLTpchKTwoFiveRrCheapCalibrationS\,s for the cheapest query per template).
Measuring the cheapest query of each template takes
\PLTpchKTwoFiveRrCheapCalibrationS\,s instead of
\PLTpchKTwoFiveRoundRobinCalibrationS\,s on TPC-H and
\PLTpcdsKOneZeroZeroRrCheapCalibrationS\,s instead of
\PLTpcdsKOneZeroZeroRoundRobinCalibrationS\,s on TPC-DS, but costs accuracy,
since short queries shift differently from long ones.

Running the round-robin queries once takes
\PLTpchKTwoFiveRoundRobinCalibrationS\,s on TPC-H and
\PLTpcdsKOneZeroZeroRoundRobinCalibrationS\,s on TPC-DS, while the
corrections above use labels from three passes. With one run per query
instead, all from one pass as in a single session, the primary method errs
by \SingleTpchWarmLowdimTemplate{} on TPC-H and
\SingleTpcdsWarmLowdimTemplate{} on TPC-DS when that run is warm, and by
\SingleTpchFirstLowdimTemplate{} and \SingleTpcdsFirstLowdimTemplate{} when
it is the query's first run in the pass; the no-move control rises about as
much (to \SingleTpchFirstControl{} and \SingleTpcdsFirstControl), so the
move still costs little.

\subsection{A neural cost model and fine-tuning}
\label{sec:neural}

\begin{figure}[t]
\centering
\includegraphics[width=\columnwidth]{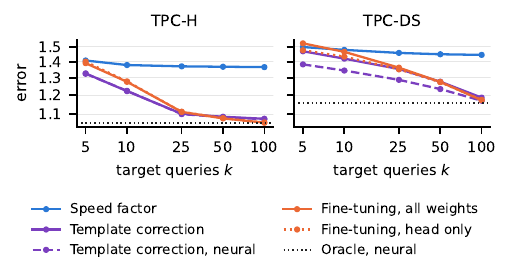}
\caption{Error against $k$ with round-robin selection, geometric mean over the 30 development pairs of each workload. Blue: the speed factor on the forest model (the cost line plus forest); solid purple: the template correction on the forest model; dashed purple: the same correction on the neural model; orange: the neural model fine-tuned on the same queries, all weights (solid) or the head only (dotted). The black dotted line is the neural model retrained on the whole target pool.}
\label{fig:nn}
\end{figure}

\begin{table}[t]
\caption{Error with round-robin selection, geometric mean over the 30 development pairs of each workload, for two source models: the cost line with a forest on its residuals, and the DACE-style neural model. Fine-tuning uses the same $k$ queries as the corrections.}
\label{tab:nn}
\small
\setlength{\tabcolsep}{3.2pt}
\begin{tabular}{@{}lrrrrrr@{}}
\toprule
& \multicolumn{3}{c}{TPC-H} & \multicolumn{3}{c}{TPC-DS} \\
\cmidrule(lr){2-4}\cmidrule(lr){5-7}
$k$ & 10 & 25 & 100 & 10 & 25 & 100 \\
\midrule
\multicolumn{7}{@{}l}{\emph{Cost line plus forest}} \\
\quad speed factor & 1.38 & 1.37 & 1.37 & 1.48 & 1.46 & 1.44 \\
\quad template correction & 1.22 & 1.10 & 1.08 & 1.42 & 1.35 & 1.19 \\
\multicolumn{7}{@{}l}{\emph{Neural (DACE-style)}} \\
\quad no adaptation & 1.84 & 1.84 & 1.84 & 1.89 & 1.89 & 1.89 \\
\quad speed factor & 1.38 & 1.37 & 1.37 & 1.40 & 1.39 & 1.38 \\
\quad template correction & 1.22 & 1.10 & 1.08 & 1.34 & 1.29 & 1.17 \\
\quad fine-tuning, all weights & 1.28 & 1.11 & 1.06 & 1.46 & 1.36 & 1.17 \\
\quad fine-tuning, head only & 1.28 & 1.11 & 1.06 & 1.43 & 1.35 & 1.17 \\
\quad oracle (whole target pool) & 1.06 & 1.06 & 1.06 & 1.16 & 1.16 & 1.16 \\
\bottomrule
\end{tabular}

\end{table}

Table~\ref{tab:nn} and Figure~\ref{fig:nn} show the neural model after the
move. Unadapted, it errs about as much as the forest model
(\NNDevTpchRrNnNoneKOneZero{} against \NNDevTpchRrHybNoneKOneZero{} on
TPC-H, \NNDevTpcdsRrNnNoneKOneZero{} against \NNDevTpcdsRrHybNoneKOneZero{}
on TPC-DS), and the speed factor helps it as much; with the speed factor or
the template correction the two models' errors on TPC-H agree to two
decimals. Fine-tuning on the measured queries is how the literature adapts
such models \cite{liang2024dace}, and it works once there are enough of them: at $k=100$ fine-tuning all
weights reaches \NNDevTpchRrNnFullKOneZeroZero{} on TPC-H and
\NNDevTpcdsRrNnFullKOneZeroZero{} on TPC-DS, against
\NNDevTpchRrHybTplKOneZeroZero{} and \NNDevTpcdsRrHybTplKOneZeroZero{} for
the template correction on the forest model. With fewer queries it falls
behind: at $k=10$ it errs by \NNDevTpchRrNnFullKOneZero{} and
\NNDevTpcdsRrNnFullKOneZero, against \NNDevTpchRrHybTplKOneZero{} and
\NNDevTpcdsRrHybTplKOneZero. Head-only fine-tuning, the cheaper variant,
gives the same errors at $k=100$. The correction also works on the neural
model: on TPC-DS at $k=100$ it reaches \NNDevTpcdsRrNnTplKOneZeroZero, level
with fine-tuning, because the neural model is the better source model there
(its oracle errs by \NNDevTpcdsRrNnOracleKOneZeroZero{} against
\PLadTpcdsOracleRandom{} for the forest model).

Per pair, the template correction on the forest model is never far behind.
At $k=100$ the 95\% interval of its error ratio to the fine-tuned network
lies below 1 in \NNDevTpchHSixFullKOneZeroZeroBelow{} and
\NNDevTpcdsHSixFullKOneZeroZeroBelow{} of 30 pairs, but its upper end is at
most \NNDevTpchHSixFullKOneZeroZeroMaxHi{} on TPC-H and
\NNDevTpcdsHSixFullKOneZeroZeroMaxHi{} on TPC-DS; at $k=10$ the ratio is
below 1 in \NNDevTpchHSixFullKOneZeroPoint{} and
\NNDevTpcdsHSixFullKOneZeroPoint{} of 30 pairs. These are development
results. On the new pairs (Sections~\ref{sec:ext} and~\ref{sec:confirm})
only H6b on SF10 holds: on JOB the neural model fits better than the
forest model, and fine-tuned it has the lower error in every pair.

\subsection{New templates}
\label{sec:unseen}

The last four columns of Table~\ref{tab:adapt} test templates absent from
the $k$ queries and from every model's training. Here no correction does
much better than the speed factor: averaged over the pairs, none of the
corrections and selection strategies we tried is more than
\XtraHardBestPct\% better, and the ridge correction is worse, since its
feature slopes, fitted to the measured templates, do not carry to a new one.
Shift models learned on another pair or on the other workload (below) do no
better: their errors are \XtraHardTransferMin{} to \XtraHardTransferMax{}
times the speed factor's. Retraining helps
little: the oracle, retrained on the target pool outside the held-out
template, errs by \PARTpchHardRandomOracleMin{} to
\PARTpchHardRandomOracleMax{} on TPC-H and \PARTpcdsHardRandomOracleMin{} to
\PARTpcdsHardRandomOracleMax{} on TPC-DS, about as much as the speed factor,
whose error is at most \PHHFourMaxAll{} times the oracle's in any pair.
Plan-level models are known to err on templates they have not seen
\cite{akdere2012}; the move to a new machine barely adds to that error.

\subsection{Larger data and another benchmark}
\label{sec:ext}

\begin{table}[t]
\caption{Error in the second round, as the geometric mean over the ordered pairs of each experiment, at $k=10$ and at $k_H$. The speed factor, affine+parallel and the ridge correction use random selection, the primary method and the fine-tuned network round robin, and 1-NN k-center. Oracle: the forest model refitted on the whole target pool; no move: the primary method on a model trained on the target itself.}
\label{tab:ext}
\small
\setlength{\tabcolsep}{3.5pt}
\centering
\begin{tabular}{@{}lrrrr@{}}
\toprule
& \multicolumn{2}{c}{E1: SF10, 6 pairs} & \multicolumn{2}{c}{E2: JOB, 20 pairs} \\
\cmidrule(lr){2-3}\cmidrule(l){4-5}
Method & $k=10$ & $k=100$ & $k=10$ & $k=33$ \\
\midrule
No adaptation & 1.66 & 1.66 & 2.71 & 2.71 \\
Speed factor & 1.31 & 1.30 & 2.51 & 2.48 \\
Affine+parallel & 1.31 & 1.26 & 2.55 & 2.42 \\
Ridge correction & 1.31 & 1.16 & 2.88 & 2.45 \\
Primary & 1.25 & 1.12 & 2.49 & 2.41 \\
1-NN & 1.33 & 1.10 & 2.65 & 3.03 \\
Network, fine-tuned & 1.27 & 1.11 & 2.16 & 2.01 \\
Network, head only & 1.28 & 1.11 & 2.08 & 2.01 \\
Oracle & 1.11 & 1.11 & 2.36 & 2.36 \\
\midrule
No move: primary & 1.15 & 1.12 & 2.43 & 2.40 \\
\bottomrule
\end{tabular}

\end{table}

The second round moves the models among the server's three environments
on TPC-H at SF10 (\XTpchTenPairs{} ordered pairs) and among five
environments of both machines on JOB (\XJobPairs{} pairs), with the first
round's methods, for $k$ up to 100 on SF10 and 50 on JOB; the registered
budget $k_H$ is 100 on SF10 and one query per template, \XJobKH, on JOB.
Table~\ref{tab:ext} gives the errors.

SF10 repeats the first round. At one query per template ($k=25$) the
primary method errs by \XeTpchTenPrimaryKTwoFive{} against
\XeTpchTenSpeedKTwoFive{} for the speed factor, near the
\XcOneTpchTenTpchKTwoFivePrimary{} and \XcOneTpchTenTpchKTwoFiveSpeed{} of
the same three environments at SF1. At $k=100$, about five queries per
template, it errs by \XTpchTenKHPrimary, against \XTpchTenKHSpeed{} for the
speed factor and \XTpchTenKHRidge{} for the ridge correction. That is level
with the no-move control (\XTpchTenKHControl) and the oracle
(\XTpchTenKHOracle). Its error is below the speed factor's in all
\XeTpchTenHTwoPoint{} pairs, though in one its interval includes 1
(Section~\ref{sec:confirm}). What changes is the small budget: at $k=10$
affine+parallel does no better than the speed factor
(\XTpchTenKOneZeroAffinePar{} and \XTpchTenKOneZeroSpeed), where at SF1 the
same environments gave \XcOneTpchTenTpchKOneZeroAffinePar{} against
\XcOneTpchTenTpchKOneZeroSpeed. \XeParallelTpchTen\% of the SF10 plans have
parallel workers, against \XeParallelTpch\% at SF1, so the line for serial
plans has almost no plans to fit, while the slowdown still differs between
templates (Section~\ref{sec:shift}), which only the offsets capture.

JOB does not repeat it. On the forest model every method errs by more than
2: at one query per template the primary method reaches \XJobKHPrimary, the
speed factor \XJobKHSpeed{} and the oracle, refitted on the whole target
pool, \XJobKHOracle. That error is the model's, not the move's: the no-move
control errs by \XJobKHControl. A template's JOB queries differ in runtime
by a factor of \XeSpreadJob{} (median geometric standard deviation, against
\XeSpreadTpch{} on TPC-H and \XeSpreadTpcds{} on TPC-DS), with
\XeJobPoolMin{} to \XeJobPoolMax{} pool queries per template to learn them
from. No source model we tried predicts them well, though some do better
than ours: refitted on the target pool, gradient-boosted trees on the same
19 features err by \XeSrcJobBoostingOracle{} and the neural model by
\NNExtJobRrNnOracleKThreeThree. Beside errors this large the shift adds
little: the speed factor's error is \XeRoomJob{} times the oracle's on
average, against \XeRoomTpch{} on TPC-H, \XeRoomTpcds{} on TPC-DS and
\XeRoomTpchTen{} on SF10, so little is left for any correction to gain.
With a lower template share, an offset taken from one query is also a
noisy estimate of its template's shift. Consistent with that, the primary
method is below the speed factor in only \XeJobHTwoPoint{} of \XJobPairs{}
pairs, and random selection does better than round robin
(\XeJobPrimaryRand{} against \XJobKHPrimary, lower in \XeJobRandBelowRr{} of
\XJobPairs{} pairs). The refits are level with the primary method (the
pooled refit \XeJobPooledRand, the forest on the errors
\XeJobForestResidRand), except target only (\XeJobTargetRand). k-center
selection, which adds the query farthest from those chosen, does worst:
with 1-NN it errs by \XJobKHKnn, more than no adaptation (\XJobKHNone),
while 1-NN on random queries errs by \XeJobKnnRand.

Fine-tuned on the \XJobKH{} queries, the neural model errs by
\XJobKHNnFull, below every method on the forest model, which decides H6
(Section~\ref{sec:confirm}). That comparison changes the source model along
with the adaptation. On the neural model itself the template correction
does as well as fine-tuning: at $k_H$ it errs by
\NNExtJobRrNnTplKThreeThree, lower than fine-tuning in
\XeJobNnTplBelowFullKH{} of \XJobPairs{} pairs (geometric mean of the ratio
\XeJobNnTplOverFullKH), and at $k=10$ lower in \XeJobNnTplBelowFullKOneZero{}
of \XJobPairs{} (\XeJobNnTplOverFullKOneZero). On SF10 the two are level at
$k=100$ (\XeTpchTenNnTplOverFullKH).

On new templates the second round repeats the first: no correction is on
average more than \XcNewTpchTenHardBestPct\% (SF10) or
\XcNewJobHardBestPct\% (JOB) better than the speed factor, and the oracle's
error is \XcNewTpchTenHardOracleMin{} to \XcNewTpchTenHardOracleMax{} and
\XcNewJobHardOracleMin{} to \XcNewJobHardOracleMax{} times the speed
factor's.

\subsection{Other source models and streams}
\label{sec:sources}

In the first round the primary method's advantage does not depend on the
source model. With a random forest, gradient-boosted trees, ridge
regression or the cost line as $f_A$, it beats the speed factor at one
query per template in all \SrcWinsTotal{} combinations of source model,
workload and pair. It also repairs weak models: on the cost line alone,
whose oracle errs by \SrcTpchCostOracle{} on TPC-H and \SrcTpcdsCostOracle{}
on TPC-DS, it reaches \SrcTpchCostPrimary{} and \SrcTpcdsCostPrimary. In the
second round it does so in \XeSrcTpchTenOtherWins{} of
\XeSrcTpchTenOtherPairs{} such combinations on SF10 but in
\XeSrcJobOtherWins{} of \XeSrcJobOtherPairs{} on JOB, \XeSrcJobCostWins{} of
them on the cost line, where the offsets lower the error from
\XeSrcJobCostSpeed{} to \XeSrcJobCostPrimary; on gradient-boosted trees, the
best of them on JOB, it errs by \XeSrcJobBoostingPrimary{} against
\XeSrcJobBoostingSpeed{} for the speed factor. We also replay each
pair's test queries as a stream in batches of about ten: it runs on the
source machine and then, without warning, on the target, and each batch is
predicted before the model learns from it \cite{gama2013}. Over the last five
batches, the primary method refitted on the last 50 queries reaches a batch
median Q-error of \StreamFinalLowdimTemplateLastFiveZero{}, against
\StreamFinalRidgeLastFiveZero{} for the ridge correction on the same window,
\StreamFinalRidgeAll{} for the ridge correction refitted on all queries
seen, whose history keeps the old machine's errors, and \StreamFinalNever{}
without updates.

A shift measured on other hardware helps a little before every template
has a query. Eight pairs of each workload have an analog, the same change
on the other machine (M1$\to$M2 and M3$\to$M4, M1$\to$M3 and M2$\to$M4, and
their reverses). A ridge model of the log slowdown on the 19 features,
fitted on the analog's pool and shifted by the mean remaining log error of
the $k$ queries, beats the primary method on TPC-H at $k=5$ in
\XtraTpchTransferWinsKFive{} of \XtraAnalogPairs{} pairs
(\XtraTpchTransferKFive{} against \XtraTpchPrimaryAnalogKFive) and at
$k=10$ in \XtraTpchTransferWinsKOneZero, but in none at $k=25$
(\XtraTpchTransferKTwoFive{} against \XtraTpchPrimaryAnalogKTwoFive). On
TPC-DS it ties at $k=5$ (\XtraTpcdsTransferKFive{} against
\XtraTpcdsPrimaryAnalogKFive) and loses from $k=10$. Part of the shift thus
carries over between machines for the same kind of change, but the
per-template offsets measured on the target soon replace it.

\subsection{Confirmatory tests}
\label{sec:confirm}

\begin{table}[t]
\caption{The five hypotheses of the first round, fixed before the labels that complete the \PHPairs{} confirmatory pairs existed (the 12 ordered TPC-DS pairs among M1 to M4 and the 6 TPC-H pairs with M4), and their results. H2 counts pairs whose 95\% interval lies below 1; H3 and H5 count pairs whose ratio is below 1, as registered; H4 bounds the speed factor's error over the oracle's.}
\label{tab:hyp}
\small
\begin{tabular}{@{}lp{0.70\columnwidth}l@{}}
\toprule
& Hypothesis and result & \\
\midrule
H1 & Template share above 0.5 in every pair (both directions of a pair share one value); smallest \PHHOneMin. & \PHHOnePass \\
H2 & $k=100$: primary (round robin) beats the speed factor (random) in every pair and the ridge correction (random) in at least two thirds; \PHHTwoVsSpeedOk{} and \PHHTwoVsRidgeOk{} of \PHHTwoN. & \PHHTwoPass \\
H3 & $k=10$: affine+parallel (random) has a lower error than the speed factor (random) in at least three quarters of the pairs that change the processor count; \PHHThreeOk{} of \PHHThreeN. & \PHHThreePass \\
H4 & New templates: the speed factor's error is at most 1.15 times the oracle's in every pair; largest \PHHFourMax. & \PHHFourPass \\
H5 & $k=100$: k-center with 1-NN has a lower error than random selection with the ridge correction in at least two thirds; \PHHFiveOk{} of \PHHFiveN. & \PHHFivePass \\
\bottomrule
\end{tabular}
\end{table}

Table~\ref{tab:hyp} lists the first round's hypotheses with their results
on the \PHPairs{} pairs that could not be evaluated when we fixed them. All
five hold under the registered rules. Counted by intervals instead, H5 still
holds (\PHHFiveOkCI{} of \PHHFiveN{} pairs), but H3 has \PHHThreeOkCI{} of
\PHHThreeN, short of its threshold: the small-budget gain of affine+parallel
is clear on TPC-H (\XtraTpchHThreeCI{} of \XtraTpchHThreeN{} pairs) and
uncertain on TPC-DS (\XtraTpcdsHThreeCI{} of \XtraTpcdsHThreeN).

The second round tests the same five hypotheses on the extension's pairs,
with the budget $k_H$ in place of $k=100$ (100 for TPC-H SF10, and one query
per template, 33, for JOB), and a sixth on the neural model. As registered,
H6 required the template correction on the forest model to beat the
fine-tuned network at $k_H$ in at least two thirds of the new pairs. On the
development pairs it did so in none at $k=100$ (Section~\ref{sec:neural}).
Before any new pair was evaluated we replaced it by two tests the
development data support. H6a: at $k=10$ the template correction's ratio to
the fine-tuned network is below 1 in more than half of the new pairs of each
workload, with a geometric mean below 1. H6b: at $k_H$ the upper end of its
interval is below 1.15 in every new pair. We report H6 as first registered
as well.

\begin{table*}[t]
\caption{The second round's tests, fixed before any of its labels existed: TPC-H at SF10 (E1, the 6 ordered pairs among M3, M4 and M6) and JOB (E2, the 20 pairs among M1, M2, M3, M4 and M6), with $k_H$ 100 on E1 and 33, one query per template, on E2. Counts as in Table~\ref{tab:hyp}; H2 gives the pairs below the speed factor and below the ridge correction, H6a the pairs whose ratio is below 1 and the geometric mean of the ratios. H6, as first registered, counts all new pairs together.}
\label{tab:hyp2}
\small
\centering
\begin{tabular}{@{}llrlrl@{}}
\toprule
& & \multicolumn{2}{c}{E1: TPC-H SF10} & \multicolumn{2}{c}{E2: JOB} \\
\cmidrule(lr){3-4}\cmidrule(l){5-6}
& Test & Count & Result & Count & Result \\
\midrule
H1 & template share $>0.5$, every pair & 6/6 & holds & 18/20 & fails \\
H2 & $k_H$: primary below speed (interval), every pair; below ridge, $\ge 2/3$ & 5/6, 4/6 & fails & 2/20, 0/20 & fails \\
H3 & $k=10$: aff.+par.\ below speed, $\ge 3/4$ of processor pairs & 2/6 & fails & 6/16 & fails \\
H4 & new templates: speed / oracle $\le 1.15$, every pair & 6/6 & holds & 20/20 & holds \\
H5 & $k_H$: k-center 1-NN below ridge, $\ge 2/3$ & 6/6 & holds & 0/20 & fails \\
H6 & $k_H$: template corr.\ below fine-tuning (interval), $\ge 2/3$ of all new pairs & \multicolumn{4}{c}{all new pairs 0/26 (E1 0/6, E2 0/20): fails} \\
H6a & $k=10$: point below 1 in $>1/2$, geo.\ mean $<1$ & 3/6, 0.989 & fails & 0/20, 1.151 & fails \\
H6b & $k_H$: upper end $<1.15$, every pair & 6/6 & holds & 0/20 & fails \\
\bottomrule
\end{tabular}

\end{table*}

Table~\ref{tab:hyp2} gives the second round, and most of its tests fail.
On SF10, H1, H4 and H5 hold. H2 fails because in one of the six pairs
(M4$\to$M6) the interval of the primary method's ratio to the speed factor
reaches \XTpchTenHTwoVsSpeedMaxHi, and H3 fails (\XTpchTenHThreeOk{} of
\XTpchTenHThreeN), likely because nearly all SF10 plans are parallel
(Section~\ref{sec:ext}). On JOB only H4 holds: the template share falls
below 0.5 in the two pairs between M1 and M3 (\XJobMOnetoMThreeShare), H2
and H3 fail with little left to correct beside the forest model's own
error, and H5 fails because k-center selection does badly there. Of the
neural tests, H6b holds on SF10 (largest upper end \XTpchTenHSixbMaxHi) and
H6a misses there with \XTpchTenHSixaPoint{} of \XTpchTenHSixaN{} pairs below
1 (it needs more than half). On JOB both fail, with the fine-tuned
network's error lower in every pair, and H6 as first registered holds in
\XAllHSixOk{} of \XAllHSixN{} new pairs. The template correction thus
carries over to ten times the data but not to JOB, where the model, not the
move, limits the error.

\section{Lessons}
\label{sec:lessons}

\textbf{Measure one query per template, chosen round robin.} On TPC-H and
TPC-DS the shift is mostly a per-template effect, so the first queries to
measure are one from each template that matters; random queries spend the
budget on large templates and leave others uncorrected. Without template
labels, k-center selection on plan features comes close there. On JOB,
where one query is a noisy estimate of its template's shift, random
selection did better and k-center worst. Avoid the cheapest query of a
template: short queries shift differently.

\textbf{Know what is left to gain.} A correction removes the shift and,
with per-template offsets, the model's bias in each template, but not its
error within a template. Where the model predicted a template's queries
well, the speed factor's error was \XeRoomTpcds{} to \XeRoomTpch{} times
the oracle's and the offsets removed most of the gap; on JOB the forest
model erred by \XJobKHOracle{} even when trained on the new machine, the
gap was \XeRoomJob, and no correction helped much. In every workload the
corrected model came on average within \XeMoveMaxPct\% of the same
correction on a model trained on the new machine, so once each template
has a measured query, what remains is that model's own error. On JOB a
better source model lowered it most: corrected the same way,
gradient-boosted trees on the same features err by
\XeSrcJobBoostingPrimary{} and the neural model by
\NNExtJobRrNnTplKThreeThree.

\textbf{Correct the best model you have; do not refit it.} An offset per
measured template on a line that treats parallel plans separately came
close to the same correction on a model trained on the new machine
(Table~\ref{tab:ladder}). On TPC-H and TPC-DS, refitting on the measured
queries, alone or pooled with the old ones, was far worse up to one query
per template, also when the refit knew the templates; on JOB the pooled
refit and the forest on the errors were level with the correction.
Fine-tuning a neural model gained at most \XtraNnGainPct\% on TPC-H and
TPC-DS; on JOB it had a lower error than the corrected forest model because
the neural model fits JOB better, and correcting the neural model did as
well as fine-tuning it.

\textbf{Expect the processor count to act through parallel plans.} When the
new machine has fewer processors than a plan's processes, plans with
parallel workers slow by \PSTpcdsMThreetoMFourParallelX{} to
\PSTpchMOnetoMTwoParallelX{} times on one processor and serial plans keep
their speed. On TPC-H a correction that knows whether a plan is parallel
lowers the error from \PATpchProcessorsRandomSpeedOneZero{} to
\PATpchProcessorsRandomAffineParallelOneZero{} with ten queries; on TPC-DS
the gain is small, and in the second round there was none: at SF10
\XeParallelTpchTen\% of the plans are parallel, and on JOB the model's own
error dominates.

\textbf{Do not expect any correction to help on new templates.} On a
template without measured queries, no correction we tried was on average
more than \XtraHardBestPct\% better than the speed factor in the first
round, or more than \XcNewTpchTenHardBestPct\% and \XcNewJobHardBestPct\%
on SF10 and JOB, and retraining did about as well as the speed factor. Its
remaining error is the model's error on a new template; to lower it,
measure a query of that template.

\textbf{Plan the measurements.} One pass over the TPC-H queries at scale
factor 1 took \EnvPassTpchMin{} to \EnvPassTpchMax{} minutes per
environment, and a label needs three; one run of the round-robin queries
took \PLTpchKTwoFiveRoundRobinCalibrationS\,s. A level shift partway
through a pass, a laptop going to sleep, other programs and a virtual
machine's clock each distorted our first attempts; Section~\ref{sec:setup}
lists the checks that caught them.

\section{Related Work}
\label{sec:related}

\textbf{Learned runtime models and hardware.} Learned models predict query
runtimes from plan features
\cite{ganapathi2009,akdere2012,marcus2019,sun2020,zhao2022}, and most are
trained for one deployment; a benchmark of seven learned cost models on one
machine found PostgreSQL's own cost model often better for the optimizer's
decisions \cite{heinrich2025}. T3 moves to new hardware by rerunning its
benchmark and retraining \cite{rieger2025t3}; zero-shot cost models
transfer across databases on one machine and leave hardware features to
future work \cite{hilprecht2022}; a recent drift benchmark and an
experimental study cover data and workload drift, not hardware
\cite{zhao2026neurbench,meng2026drift}; and a recent tutorial lists aligning
label collection with the production hardware as open
\cite{heinrich2025tutorial}. Redshift's Stage predictor feeds the instance
type to a global model trained across its fleet and backs it with a model
and a cache of repeated queries for each instance \cite{wu2024stage};
Costream takes hardware features as input for stream processing
\cite{heinrich2024costream}; and Packing Light predicts the throughput of
TPC-H and TPC-DS mixes on new cloud hardware from fingerprints taken on a
few sampled configurations \cite{duggan2013}. The closest work moves a cost
model to a second machine: QCFE refits per-operator coefficients from
queries run there and retrains on 200 labeled queries \cite{yan2024qcfe},
DACE fine-tunes on a whole workload rerun on the second machine
\cite{liang2024dace}, and ParamTree predicts PostgreSQL's cost constants
from a machine's hardware and settings and, on unseen cloud servers, beats
both a linearly scaled cost model and one calibrated on samples from each
server \cite{yang2023}. Older work calibrates cost formulas from measured
queries: Wu et al.\ calibrate PostgreSQL's cost units on each machine with
designed queries \cite{wu2013}, and Zhu and Larson fit a regression cost
formula per query class of a multidatabase system from queries sampled in
that class \cite{zhu1994}, a precursor of our per-template offsets, as is
Cleo's combination of many small cost models for recurring subgraphs with
general ones \cite{siddiqui2020cleo}; Kepler likewise learns per
parameterized query, choosing among candidate plans from their measured
runtimes \cite{doshi2023kepler}. We add a controlled measurement of how
much of a hardware shift a handful of queries can absorb, and of which
queries to run, and we compare against fine-tuning a DACE-style model on the
same queries.

\textbf{Transferring performance models.} For configurable software, a
fitted line from source to target performance (model shift) carries a model
across hardware when the change is mild \cite{valov2017,jamshidi2017}, and
fails for multithreaded programs or severe changes
\cite{valov2020,jamshidi2017,jamshidi2018}; Libra, which moves database
configuration models across hardware and workloads, finds that model shift
can transfer negatively \cite{nakamori2026libra}. OneProxy adapts a
neural-network latency predictor to a new device from about 50 measurements
with one global factor plus sparse per-operator deviations and aims to
preserve rankings \cite{lu2021oneproxy}; our ridge correction has the same
form with an L2 penalty. The affine part of our affine+parallel correction
is model shift on the log prediction, and our results show where it stops:
a change of processor count acts through parallel plans, and in our first
round every change acted through templates.

\textbf{Model updating and choosing samples.} Clinical prediction research
calls the speed factor \emph{recalibration in the large} and finds that such
simple updates beat revising the whole model unless the new sample is large
\cite{steyerberg2004,vergouwe2017}. Our ridge correction is regularized
least-squares transfer on the source model's residuals, with an unpenalized
intercept and a leave-one-out penalty \cite{kuzborskij2013}. To choose
queries we use greedy D-optimal design \cite{wynn1970,mitchell1974} and
farthest-first selection \cite{gonzalez1985,sener2018}. Ernest picks the
training runs of its linear runtime model by A-optimal design under a budget
on their cost \cite{venkataraman2016ernest}, which would counter the long
queries that D-optimal selection picks here; L2S uses the source model to
choose which configurations to measure \cite{jamshidi2018}, Warper which
queries to use when it adapts a cardinality estimator \cite{li2022warper},
and Ma et al.\ which queries to label for learned database components
\cite{ma2020active}. Learned database models have also been updated for
shifts in data and workload \cite{kurmanji2023,wui2024}.

\section{Threats to Validity}
\label{sec:threats}

\textbf{External validity.} The study covers one database system. The SF1
data fit in memory in every environment, and the extension adds data ten
times larger and a benchmark on real data, but not data larger than memory.
It uses two physical x86 machines, both virtualized; a third experiment,
on a machine that is not, was registered in case one became available,
and was not run. The server's guest sees no frequency scaling and no SMT
siblings, and Windows schedules the laptop's virtual processors on its
performance and efficiency cores alike, so pinning there fixes a virtual
processor, not a kind of core; no pinned environment is noisier than the
noisiest of the others (Table~\ref{tab:env}). Queries ran warm and one at a time,
and only TPC-DS templates whose first query ran within 30\,s on M1, and
JOB queries under 60\,s, entered. The
processor changes cut a plan's leader and workers to one or two processors,
and pinning keeps the planner's two workers per Gather; a target with more
processors than a plan's processes would hide most of that effect. Plans
are fixed, so a move that also changes settings or plans is out of scope.

\textbf{Internal validity.} The pairs of a workload reuse the labels of its
environments (30 pairs of six in the first round, \XTpchTenPairs{} of
three and \XJobPairs{} of five in the second), so they are not independent
replications; we report intervals per pair and count pairs instead of
pooling them. JOB's test set holds about one query per template, so its
intervals are wide, though its failed tests fail on the point estimates
alone as well. We designed the
methods on TPC-H pairs among M1 to M3 and confirmed them on pairs measured
later. The server is shared, so its passes started at quiet hours and every
pass was checked for level shifts. Per-node timing adds overhead that
differs between machines, so the operator analysis is qualitative.

\textbf{Construct validity.} We score per-template medians and report
tails and bias only at one query per template
(Section~\ref{sec:adapt}). All source models except the neural one
use the same 19 plan-level features, and the neural model is a
reimplementation of DACE's architecture trained on root runtimes only, not
DACE's pretrained model.

\section{Conclusion}
\label{sec:conclusion}
In our environments, moving a runtime model to new hardware did not need a
new training set. In all four workloads, one measured query per template,
with an offset for each template on the moved model, came on average within
\XeMoveMaxPct\% of the same correction on a model trained on the new
machine; separate lines for serial and parallel plans mattered mostly
before every template had a query. Whether the corrected model is accurate
depends on the model. On TPC-H at both scales and on TPC-DS it came close
to the oracle. On JOB the correction removed the shift but the error
stayed high: the forest model erred by more than a factor of two even when
trained on the new machine, boosted trees and a neural model fit JOB
better, and the neural model corrected the same way did as well as
fine-tuned. JOB's shift also lay less between templates, below half for
the move between M1 and M3, and most of our registered tests failed there.
For templates never measured, no correction we tried was on average much
better than the speed factor in either round, because the remaining error
is the model's error on a template it has not seen, and the change of
machine adds little to it. Operator-level corrections, other database
systems, plans that change with the machine and data larger than memory
are the next steps.

\section*{Acknowledgment}
The author thanks Yao-Liang Yu for feedback on the project proposal from
which this study grew.
Claude (Anthropic) was used in sections III and V of this paper to draft and organize the text, tables and figures, and to write and run the code for the experiments. The author directed the work and is responsible for all of its content.

\bibliographystyle{IEEEtran}
\bibliography{references}

\end{document}